# DeepMech: A Machine Learning Framework for Chemical Reaction Mechanism Prediction

Manajit Das[a], Ajnabiul Hoque[a], Mayank Baranwal[b,c,*], and Raghavan B. Sunoj[a,d,*]

[a] Department of Chemistry, Indian Institute of Technology Bombay, Powai, Mumbai 400076, India.

[b] Department of Systems and Control Engineering, Indian Institute of Technology Bombay, Powai, Mumbai 400076, India.

[c] Tata Consultancy Services Research, Mumbai, India.

[d] Centre for Machine Intelligence and Data Science, Indian Institute of Technology Bombay, Powai, Mumbai 400076, India.

**Abstract**

The ability to predict the complete, step-by-step mechanism of chemical reactions from first principles remains a grand challenge in science. Importance of chemical reaction mechanisms (CRMs) pervades almost all domains such as prebiotic chemistry, drug discovery, materials science and so on. Uncovering CRMs remains a complex task, traditionally reliant on expert-driven experiments or expensive quantum chemical computations. While deep learning (DL) studies have shown promise in predicting reaction outcomes, it largely ignored important intermediates and mechanistic steps *en route* to the product of immediate interest. Since sequence-to-sequence models that generate products character by character are prone to hallucination, we consider it important to develop DL models that prioritize reactivity over learning the syntax and semantics of sequences. The progress in reaction mechanism predictions is further limited by the lack of large-scale, mass-balanced datasets with mechanistic annotations.

Motivated by these key lacunae, we introduce DeepMech, an interpretable graph-based DL framework that employs attention mechanisms at both atom and bond levels. Instead of end-to-end learning from reactants to products, our model incorporates a template of mechanistic operation (TMOp) for the generation of intermediates in elementary mechanistic steps. It leverages TMOps, to predict step-by-step mechanisms toward realizing full CRMs for a multitude of reaction classes of high contemporary significance. To train our DeepMech model, first we construct ReactMech, a meticulously curated dataset of about 30K full reaction mechanisms, each comprising of several atom-mapped elementary steps (totaling to 100K). DeepMech achieves the state-of-the-art accuracy of 98.98±0.12% in predicting elementary steps and 95.94±0.21% in complete CRM tasks. The model maintains high fidelity even in out-of-distribution scenarios involving unseen catalysts, ligands, and/or mechanistic classes. In so far as the generalizability goes, DeepMech effectively reconstructs multistep CRMs relevant to prebiotic chemistry, beginning from trivial primordial substrates such as nitrogen, ammonia, methane, water, and hydrogen cyanide, to complex biomolecules like serine and aldopentose. Further, attention-based interpretability analysis reveals that DeepMech correctly identifies reactive atoms and bonds, in line with chemical intuition. Collectively, DeepMech offers a promising step toward data-driven prediction of CRMs, with potential to expedite mechanistic understanding and reaction design across domains.

## Introduction

Chemical reactions form the foundation of synthesis of vital compounds such as drugs, pharmaceuticals, agrochemicals, etc., enabling the construction of complex molecules bearing tailored properties.[1] In a chemical reaction, reactants convert to products through one or more elementary steps (Figure 1a). A meaningful description of such elementary steps and the associated intermediates in a reaction is referred to as reaction mechanism. A deeper understanding of chemical reaction mechanism (CRM) is indispensable in discovering new reactions and in designing molecules of high importance.[2] Heuristic methods, such as the arrow-pushing formalism, have traditionally been used to rationalize reaction pathways as well as in generating new mechanistic hypotheses.[3] Although chemical intuition is generally very valuable, it could become inadequate for navigating newer and more complex reaction systems. Translating these heuristics into a computationally tractable framework and the development of data-driven models remains challenging. Therefore, the development of reliable, robust, and efficient computational methods for CRM prediction is of great current interest.

Several methods have been developed to design complete CRMs.[4] For instance, quantum chemical methods, although tedious; can offer valuable mechanistic insights (Figure 1b).[5] Second, automated graph-based sampling can enable the exploration of various reaction pathways, which are then refined using techniques such as the Nudged Elastic Band algorithm.[6] Despite their effectiveness, these approaches are limited by their high computational costs and are often confined to applications to small molecules (typically fewer than 100 atoms). Given the significance of CRM in a broad array of reactions of high practical value, development of reliable ML models for CRM prediction can be regarded important.[7] Such models can help bridge the gap between heuristic approaches and automated learning. However, it would inevitably require the development of advanced algorithms and access to large, mass-balanced, and chemically plausible datasets to train the ML models. It is expected that a well-developed ML model should be competently able to match expert-level reasoning in CRM prediction tasks. Perusal of contemporary literature indicates that the data-driven ML models have achieved impressive success in predicting major products in reactions, often abstracting chemical complexity to simplify learning.[8] Models trained on large-scale experimental datasets have excelled in tasks such as SMILES translation and classification of reaction templates[9] However, most of the previous efforts focused only on the final products, neglecting mechanistic details such as the movement of electron, identities of key intermediates, invoking catalyst participation, etc.

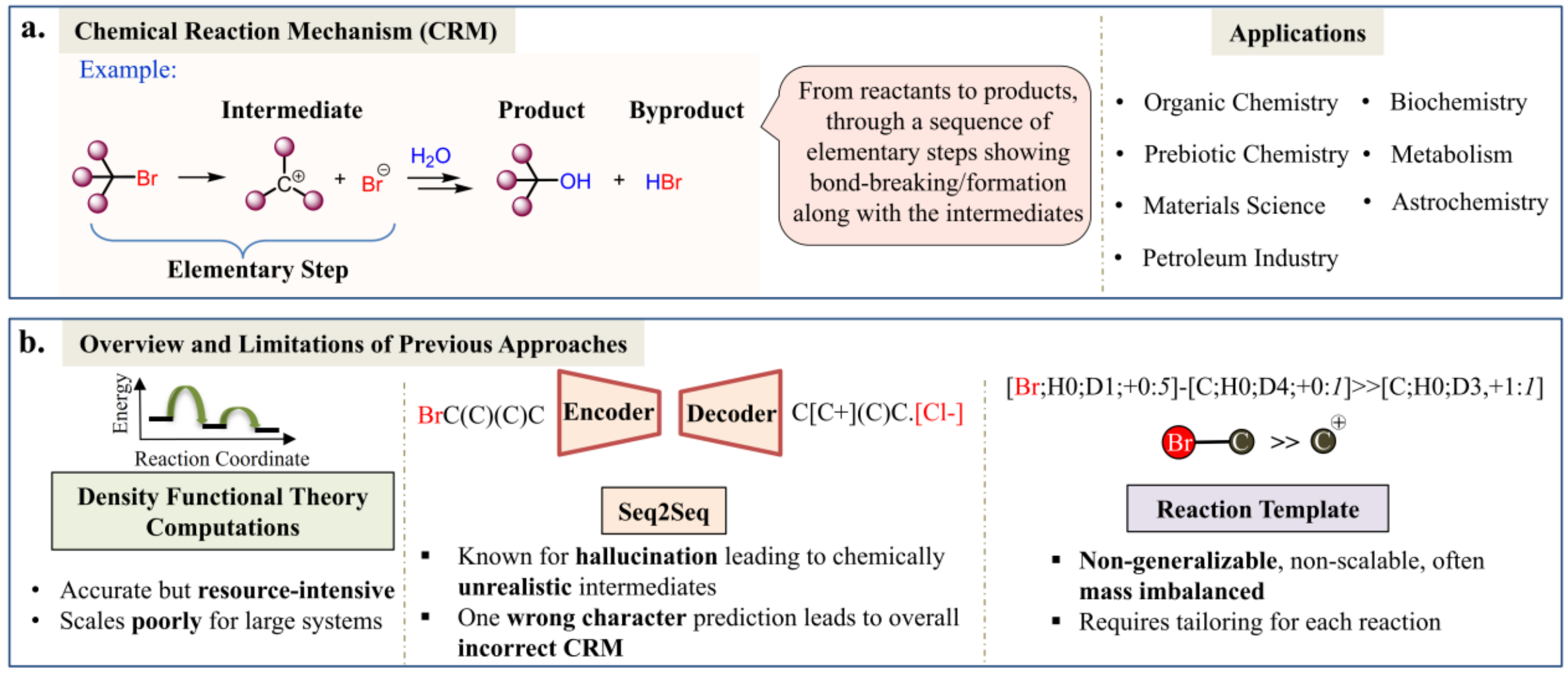

**Figure 1.** (a) A representative elementary step involving a heterolytic bond breaking resulting in a carbocation in an $S_N1$ reaction mechanism. (b) An overview of previous approaches used in reaction mechanism prediction and their key limitations.

Recent advances in integrating mechanistic details into chemical reaction datasets have led to some progress, but limitations remain. Although large reaction corpora like USPTO-Full[10] and Pistachio provide extensive coverage of chemical reactions, they do not include mechanistic annotations. Graph-based generative models, such as ELECTRO, attempt to infer electron flow between reactants and products without explicitly considering reactive intermediates.[11] Timely efforts by Baldi and coworkers have laid an excellent foundation for data-driven prediction of CRMs. Their early works used the atom-level features to identify the electron-donating (source) and electron-accepting (sink) atoms, thereby enabling an automated generation of possible mechanisms while implicitly maintaining mass balance through source–sink pairing.[12] Their subsequent work used a deep learning architecture, such as the LSTM, to enhance the accuracy of source and sink identification in about 11K elementary steps.[13] More recently, they employed a hybrid ensemble approach by combining Chemformer models with a rule-based filtering to ensure mass- and charge-balanced products on larger 13K elementary steps from PMechDB dataset.[14] More importantly, mechanistic interpretability in the form of arrow-pushing diagrams was made possible by using the ArrowFinder platform.[15] Building on a similar conceptual foundation, the Jung group proposed Reactron, a GNN-based model built on a labeled dataset of 2.8M elementary steps (excluding radical and transition metal catalyzed reactions) to predict reactive sites.[16] They have previously developed MechFinder using a subset of 31K reactions curated from the USPTO dataset to obtain chemically reasonable arrow pushing diagrams.[17] However, MechFinder does not consider a large family of reactions belonging to transition metal catalysis. The associated dataset includes only the source and sink information, omitting intermediate structures and elementary steps. Given that a plethora of industrially important reactions employ transition metal catalysts, it is desirable to develop ML models suitable for such settings as well.

Coley and coworkers advanced this area by adapting end-to-end reaction prediction models to elementary steps, creating Graph2SMILES (G2S) based on the Pistachio database.[18] However, their curated dataset is not fully mass balanced and does not explicitly account for proton sources/sinks, leading to errors in situations such as in the prediction of alcohol deprotonation under base free conditions. In the most recent innovation, they have employed a diffusion-based generative framework to model electron flow as a continuous flow-matching dynamics problem.[19] The FlowER model integrates the graph transformer architecture with explicit mass balance over 1.6M elementary steps to achieve excellent performance in CRM predictions. A summary of the major ML approaches for CRM prediction is provided in Table 1 to provide a contextual background and relations to our model presented later in the manuscript.

**Table 1.** Comparison of Representative ML Models for CRM Prediction

| details of | Baldi (2011-2025) (refs.12-15) | Jung (2024) (ref.16) | Coley (2024-2025) (refs.18a and 19) |
|---|---|---|---|
| model | ANN, LSTM, Chemformer | GNN | G2S, FlowER |
| innovation | source-sink prediction | source-sink prediction | flow matching for mass-balanced generation |
| scope | polar, radical, pericyclic reactions | polar (no radical and transition metals) | broad range of organic reactions |
| dataset size | 13K | 2.85M | 1.6M |
| mass-balance | yes, implicit through source-sink prediction | implicit | G2S (no), FlowER (yes) |
| interpretability | allows visualization of arrow pushing | limited | limited |

Despite the burgeoning recent efforts in using deep learning for CRM prediction, there remains significant room for improvement toward achieving accurate, scalable, and generalizable predictions. Existing sequence-to-sequence models are prone to hallucinations, often resulting in mass-imbalanced products. While source–sink prediction frameworks maintain mass balance besides offering arrow-pushing representations for a majority of elementary steps, some of critical steps such as oxidative addition and reductive elimination typically found in transition metal catalysis are not well described using the arrow-pushing approach.[20] Although electron-flow-based models are conceptually appealing, they offer only limited interpretability. Further, the training would require explicitly hydrogen-mapped elementary steps and their downstream use might as well warrant fine-tuning to impart model with broader generalization capabilities. To address some of these limitations and to complement the existing approaches, we propose a mass-balanced, template-based, attention-guided graph neural model designed to: (a) minimize hallucination by avoiding direct SMILES generation, (b) provide interpretability through attention visualization, and (c) effectively model transition metal-catalyzed reactions, where conventional arrow-pushing are not straightforward but explicit bond-breaking and bond-forming events can still be captured. We set the following as the major objectives of this work:

- To create a reaction mechanism dataset ReactMech, that complements reaction databases like USPTO by providing mass-balanced elementary step mechanisms with explicit intermediates. ReactMech provides complete reaction pathways (reactants → intermediates → products), including transition metal catalyzed coupling reactions and C–H bond activation reactions, with precise atom mapping to enable mechanistic interpretability.
- To develop a robust and interpretable ML model for CRM prediction: We develop DeepMech, the first model capable of predicting the complete reaction mechanism by learning template of mechanistic operations (TMOp). The model is desired to have (a) a strong generalization performance, in both in-distribution (ID) and out-of-distribution (OOD) reactions, (b) ability to identify side products and/or byproducts, which are important in reaction development, and (c) chemical interpretability for the predicted mechanisms.
- To explore prebiotic mechanisms: We extend our approach to prebiotic chemistry by curating a novel dataset, PrebioMech consisting of mechanistic pathways. Our DeepMech predicts plausible pathways toward higher-order amino acids and sugars, which are fundamental building blocks, respectively for proteins and carbohydrates in living systems, thus contribute toward understanding of chemical evolution in prebiotic environments.

## Methods

We have organized the methods section in four major subsections to provide a concise detail of each of the distinct, but inter-dependent, building blocks in our workflow. First, steps involved in creating a multi-step reaction mechanism dataset (ReactMech) are described. Next, the key components in building our graph-based DL model (DeepMech) are presented. Subsequently, we describe how the prediction of reactive bonds and TMOps by DeepMech are used to infer the intermediate or product structures in elementary steps of chemical reactions. The utilization of the beam search algorithm for ranking top-*k* predicted mechanistic pathways is explained before beginning discussions on our results.

### A. The Dataset: ReactMech

In this study, we use a subset of the reaction data extracted from the USPTO as compiled by Lowe.[10] Given that this subset does not contain any transition metal catalyzed reactions, we augment our dataset with 7 additional reaction classes to broaden the mechanistic coverage. Thus, our ReactMech dataset consists of 60 different reaction classes (see Section 2 in Supporting Information) from the USPTO and 6 representative sets of transition metal catalyzed coupling reactions (Buchwald–Hartwig amination (BHA),[21] Suzuki–Miyaura,[22] Kumada,[23] Heck,[24] relay-Heck,[25] and C–H activation reactions).[26] In addition, the inclusion of Diels–Alder cycloaddition reactions[27] brings a desirable mechanistic diversity in the form of pericyclic reactions. Collectively, the ReactMech dataset provide 29,604 reactions belonging to 67 different reaction classes.[28] We utilized this dataset to create atom-mapped and mass-balanced full CRMs, providing a total of 104,964 elementary steps. To our knowledge, such a comprehensive mechanistic dataset is currently absent in literature that encompasses a broad range of reaction classes characterized by their underlying mechanisms.

We illustrate the key steps involved in the data preparation using Figure 2. A simple reaction (Rxn:1, row-1) between acetyl chloride and isopropylamine leading to *N*-isopropylacetamide as the product is considered here. This reaction belongs to a widely found class of addition-elimination reaction exhibited by carbonyl derivatives. The domain knowledge and literature reports can help propose a plausible CRM, consisting of one or more elementary steps with relevant intermediates {$\mathbf{I_1}$, $\mathbf{I_2}$, …,$\mathbf{I_n}$} with full mass balance at each such step (row-2). In every step, a few of the atoms undergo changes in their environment (e.g., charge, number of hydrogen atom, bonding). It is important to map the atoms of the reactants precisely to corresponding atoms in the products. Note that the reactants and products here also encompass the intermediates *en route* to the product. Each of these atom mapped elementary step, leading to intermediates or products, constitutes an instance in the ReactMech dataset.

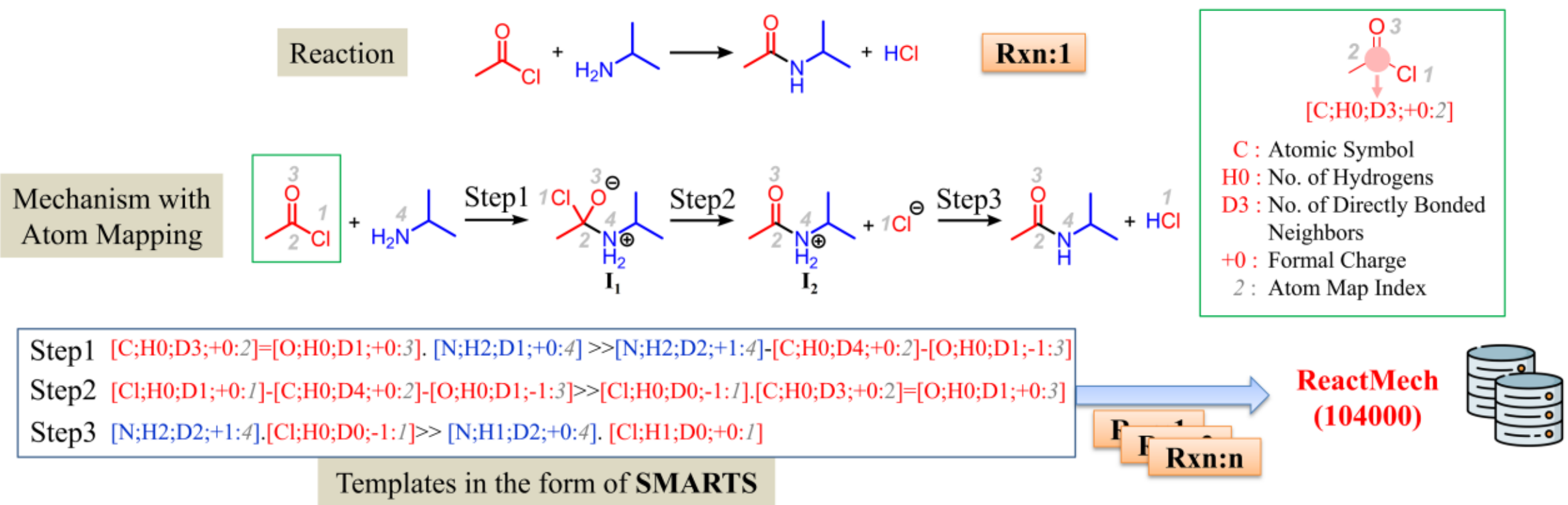

**Figure 2.** An overview of the stepwise chemical reaction mechanism (CRM) generation to form ReactMech dataset for elementary steps. A chosen reaction (e.g., Rxn:1) is first subjected to annotation of mechanism and atom mapping. For each elementary step (e.g. Step1, Step2, and Step3) mechanistic templates in the form of SMARTS are extracted. These templates are then used to generate CRMs for other reactions (Rxn:2,…,Rxn:n) belonging to the same class. The resulting set of elementary steps collectively forms the ReactMech dataset. Shown in the inset is an example of a SMARTS pattern for a carbon atom in one of the substrates in Rxn:1.

Next, we developed an algorithm to automate the generation of atom mapped elementary steps for all the other reactions in the same class. From each atom mapped elementary step, the corresponding SMART-based template with full specification of the reactive atoms and bonds, is created as shown in row-3 of Figure 2. The algorithm takes the initial reactant(s) and then applies a sequence of SMARTs-based templates iteratively i.e., the intermediates generated by a previous template serve as the input for the next template until the final product is obtained. Similarly, the remaining reactions (e.g. Rxn:2,...,Rxn:n) are passed through these templates to derive all the atom mapped elementary steps. For the reactions where the reactants couldn't be matched with any of the preset templates, we consider it as a failure mode and tackle them individually as follows. A template mismatch reaction is first subjected to a chemically meaningful mechanistic annotation (row-2, Figure 2), followed by atom mapping and then a new/modified template suitable for that reaction class is generated (See Section 7 in the Supporting Information). This approach would readily facilitate inclusion of new reaction classes on which our algorithm can be applied to generate the corresponding reaction mechanism. Such modular operational advantage helps in expanding the scope of the ReactMech dataset for new reaction classes. The ReactMech, with 67 reaction classes and 29,604 reactions in it, gives a total of 104,964 elementary steps, providing a wide coverage of different chemical reactions.[29]

**B. The Model: DeepMech**

The details of the DeepMech workflow for predicting elementary steps as well as full CRM are provided here.[30] As shown in Figure 3a, the model operates on molecular graphs ($G$), which serve as a structured representation of the chemical species involved. Each graph encodes atoms as nodes ($V$) and bonds as edges ($E$), with initial atomic features $h_u^0$ derived from the Weave featurization strategy.[31] A Message Passing Neural Network (MPNN) is employed to update atomic features $h_u$, where each atom $u$ receives messages from its neighbors $v$.[32] Through several message-passing steps, the atomic embeddings are iteratively refined to capture local structural information. Next, we apply a distance-aware Global Reactivity Attention (GRA) module, based on multi-head self-attention, on the MPNN-derived atomic embeddings to capture long-range dependencies.[33] This module integrates an Atom Distance Matrix (ADM), where each element $r_{u,v}$ denotes the topological distance (i.e., number of bonds) between atom pairs $u$ and $v$.[34] By incorporating these distance-aware relationships, the GRA module selectively attends to chemically relevant atom pairs, such as those involved in rings, aromatic systems, or functional groups, thereby enhancing the identification of potentially reactive sites.

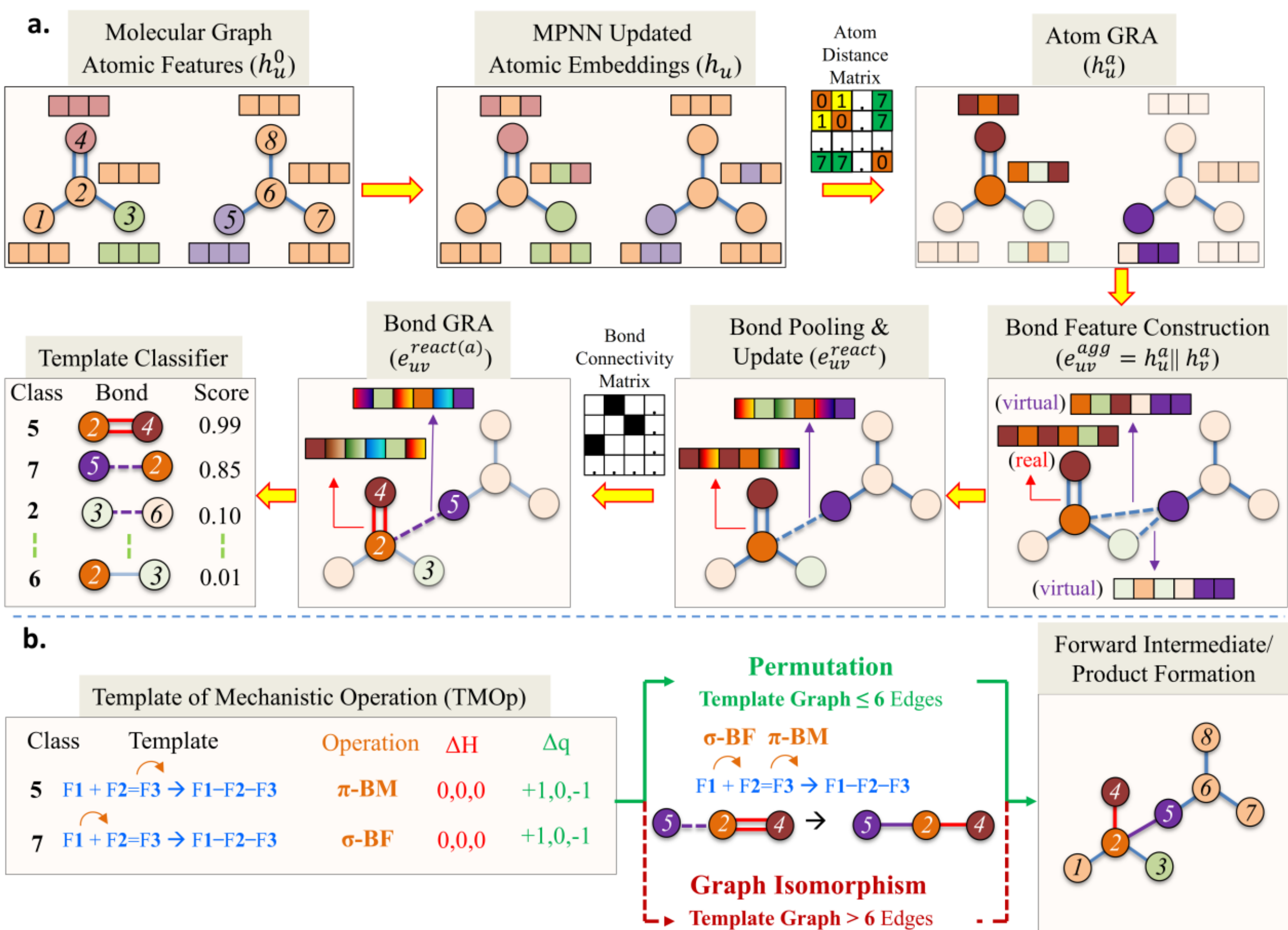

**Figure 3.** Overview of the DeepMech workflow for prediction of elementary steps in chemical

reaction mechanism. (a) In top right corner, the atoms with higher attention are shown with a deeper color while those with a lesser attention appear faded. In the second row, rightmost box shows a subset of real and virtual bonds for improved clarity. In the box denoted as Bond GRA (global reactivity attention), those bonds receiving higher attention are given different colors (red and violet for real and virtual bonds respectively) with a higher intensity, while those with lesser attention are shown faded. (b) In the last row an illustration of TMOp is given, where ΔH and Δq respectively denote the changes in the number of hydrogen atoms and formal charges. The actions such as σ bond formation (σ-BF), σ bond breaking (σ-BB), and π bond modification (π-BM) involved in TMOp are shown.

For each atom pair ($u$,$v$), we construct a bond-level feature vector $e_{uv}^{agg}$ by concatenating their refined atomic embeddings as shown in Figure 3a (second row). This is done for both the real bonds, i.e., the edges present in the molecular graph, and the virtual bonds connecting atom pairs that are not directly bonded.[35] The virtual bonds are shown using dashed lines in Figure 3a. These virtual connections are designed to capture potential bond formation that is expected to occur during a chemical reaction, which facilitate the model to learn intermolecular interactions. To identify the top-$k$ most reactive bonds $e_{uv}^{react}$, we apply a reactivity pooling strategy. Further, to account for the structural context and relational dependencies among the reactive bonds, an additional GRA module is introduced. This module operates on a Bond Connectivity Matrix, which encodes whether two bonds share a common atom, thereby indicating direct topological relationships within the molecular graph. For each pooled reactive bond, a multiclass classifier is trained to predict the probability distribution over a predefined set of 545 TMOps, which consider the changes in charge and number of hydrogen atoms. Each TMOp encodes three key components, such as (i) a generalized template (hereafter referred to as the 'template'), (ii) nature of the mechanistic operation, and (iii) the change in hydrogen count and formal charges as shown in Figure 3b.

It should be noted that a template carries the details of mechanistic changes and is devoid of atom specific properties (i.e., atom type, number of hydrogen atoms attached, charge).[36] A comparison between Figures 2 and 3 brings out the advantage of our approach. The atom-specific template for Step 1 shown in the third row in Figure 2 is only applicable to a nucleophilic addition of any primary amine on a carbonyl carbon. On the other hand, a template such as F**1**.F**2**=F**3**>>F**1**−F**2**−F**3**, can represent a general bond formation between F**1** and F**2** and a double bond breaking between F**2** and F**3**. Therefore it can be applied to any other reaction with this kind of elementary step, such as a nucleophilic addition of an amine, alcohol, or thiol to double bonded subsystems such as a C=O in aldehydes, ketones, or other carbonyl derivatives encompassing a large group of reactions. Similarly, the subsystem could as well be C=C in alkenes, C=N in imines, and S=O in sulfonyl derivatives. This approach therefore provides significant advantages over the atom specific template-based methods, which rely on atom-specific constraints. Next, key feature of TMOp is the inclusion of the operation type such as a σ bond formation (σ-BF), denoting the flow of electrons from an electron rich to electron poor atom/center, σ bond break (σ-BB), π bond modification (π-BM), and hydrogen atom exchange (HAX). Furthermore, the changes in the hydrogen count and formal charge are inherently taken into account in TMOp. Figure 3b illustrates a TMOp consisting of two a σ-BF (bond formation between F**1** and F**2**) and a π-BM (conversion of a double bond to a single bond between F**2** and F**3**) (see Section 3 in Supporting Information for details). The design of TMOp was inspired, in part, by the Generalized Reaction Template (GRT) framework introduced by Jung et al.[35] However, unlike GRT, which was developed primarily for reaction template extraction and generation, TMOp is specifically meant for stepwise mechanistic prediction, enforcing mass balance and explicit atom participation in every elementary step. Additional discussion highlighting these distinctions, including treatment of lone-atom reactivity and those involving aryl rings, is provided in Section 5.7 of the Supporting Information. After the classification, TMOps and their associated pooled bonds are ranked by the corresponding probabilities, and the top-ranked ones are sequentially selected until their combined operations satisfy the corresponding TMOp. This TMOp is then applied on the identified reaction centers to generate the predicted structure, as detailed in the following section.

**C. TMOp (Template of Mechanistic Operation) to Predicted Structure**

To fit the predicted bond sets with the predicted TMOps, we implement a strategy based on the size of the template graph present within the TMOp. For template graph of ≤6 edges, we consider the full permutation space of the *n* predicted bonds to balance tractability and coverage. This permutation-based matching can be understood with the help of an illustrative example shown in Figure 3b where the template graph has (**1**,**2**) and (**2**,**3**) as its two edges (Figure 3b). Note that the numbers in bold font type denote numbering used in the template, while italic font represents the atom index in the predicted bonds such as (*2*,*4*), (*3*,*6*), and (*2*,*5*). Here, the task is to identify valid atom to template mapping that preserve the edge connectivity of the template. Since the predicted bonds are three and the template edges are two, we evaluate all $^{3}P_{2} = 6$ permutations of the predicted bonds. These permutations can be aligned with the ordered nodes from the template graph. For example, under the mapping {**1**→*5*, **2**→*2*, **3**→*4*}, the edges (**1**,**2**) and (**2**,**3**) get correctly mapped to the predicted bonds (*5*,*2*) and (*2*,*4*) respectively, which corresponds to one of the permutations (middle inset in Figure 3b). Here, **1** refers to F**1** in the template and *5* denote the predicted attentive atom. Similarly, mapping {**2**→*2*, **3**→*4*} respectively pertains to F**2** and atom *2* and F**3** and atom *4*. Thus, this permutation is considered a valid match, which will lead to the predicted structure. Conversely, an alternative permutation involving (*2*,*4*) and (*3*,*6*) as the predicted bonds will not lead to any valid mapping, rendering such permutations invalid as it fails to preserve connectivity.

While the above-mentioned approach can systematically explore the permutation space to identify structurally consistent mappings with small size graphs, it can become computationally intractable for larger graphs. Even in the case of a system with 12 predicted bonds and a template graph with $r$ = 7, 8, or 9 edges, the number of possible permutations become as large as 4M, 20M, and 80M, respectively. Thus, for graphs with more than 6 edges, we leverage subgraph isomorphism techniques to avoid the combinatorial explosion of permutations. The idea is to identify isomorphic subgraphs of the true structure from the predicted bonds.[37] Establishing these correct mappings between the predicted bonds and the template enables the derivation of the forward intermediate or product structure. Although this approach can predict the immediate next intermediate relevant to elementary step prediction, it does not provide the complete CRM. To achieve our objective of CRM prediction, we therefore implement a beam search algorithm designed to efficiently explore and rank multiple plausible mechanistic pathways.

**D. Beam Search**

DeepMech generates top-*k* predictions for a given input to construct a complete CRM using a tailored beam search algorithm in conjunction with a Reaction Classifier as shown in Figure 4.

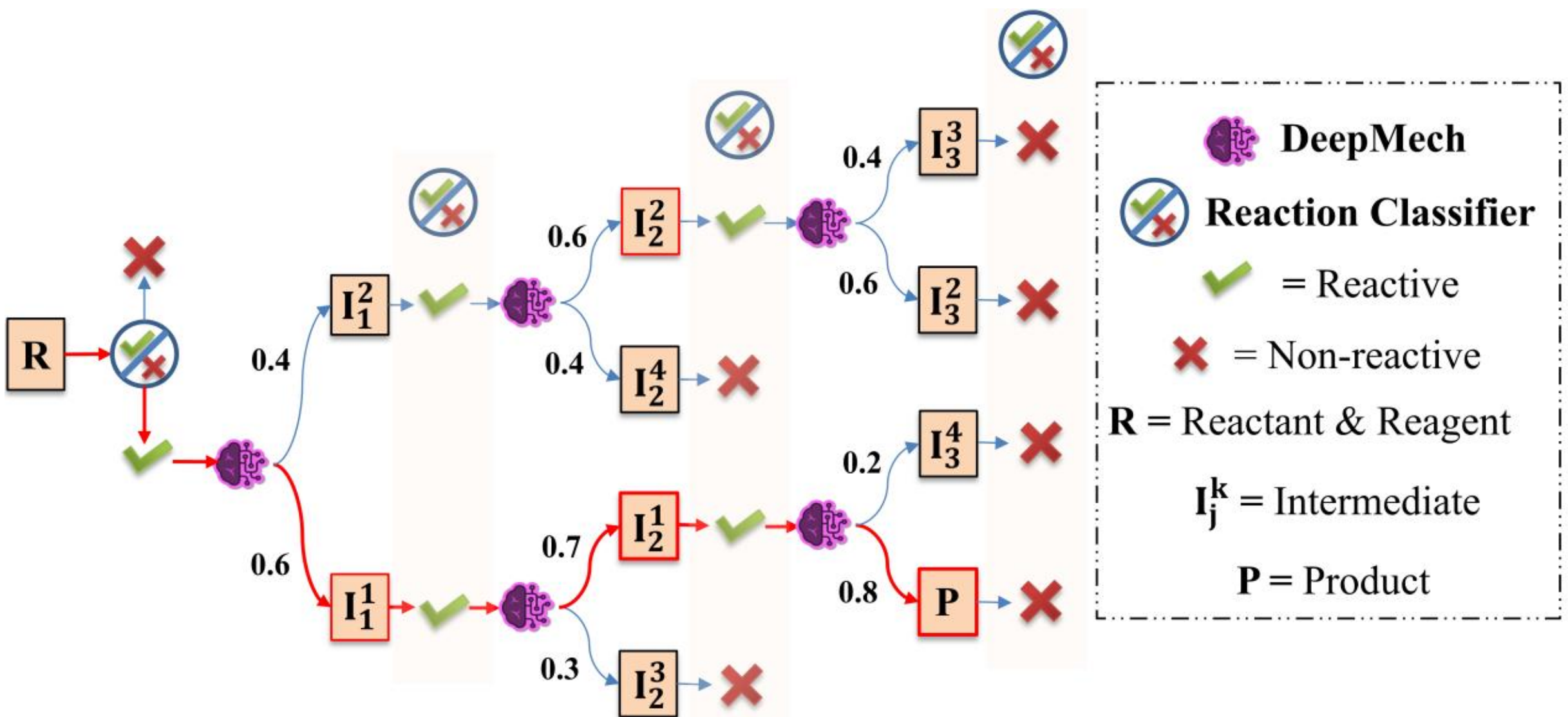

**Figure 4.** Illustration of complete CRM prediction using the beam search algorithm.

Given a set of initial chemical entities (i.e., reactants, catalyst, and other reagents) as the input to the model, it is first evaluated by the AttentiveFP-based Reaction Classifier to verify whether a reaction is likely or not.[38] The classifier learns to recognize chemically valid, yet non-reactive, molecular combinations from the training instances derived from stable reaction products. Consequently, it distinguishes chemically plausible assemblies on the basis of their reactivity, not just their structural validity. It is also worth noting that the model was never exposed to invalid or non-physical species, thus ensures that reactivity predictions reflect true chemical feasibility (see Section 5 in the Supporting Information for more details). This classifier, which is trained to label the input as reactive or non-reactive, serves as a stopping criterion for the CRM generation if a given input is predicted as non-reactive, leading to a termination of the process at this very step. Otherwise, the model proceeds to propose the top-*k* most probable next products, along with their associated probabilities. While DeepMech allows mass-balance via TMOp, it is not enforced as a hard constraint. It is therefore likely to experience some rare situations of mass imbalance due to hydrogen/proton. To circumvent this, we apply domain-specific chemical heuristics to the top-*k* predictions to filter out chemically implausible products and to prioritize among predictions with comparable probabilities (see Supporting Information in Section 5 for more details). Each product is then evaluated by the same Reaction Classifier to determine whether the predicted CRMs have reached completion (i.e., Reaction Classifier predicts it as non-reactive) or it should be further expanded. The CRMs deemed incomplete are expanded through additional predictions by the DeepMech model, with their cumulative log probabilities updated at each step. This process continues until all candidate CRMs are classified as complete or a predefined maximum depth (which is set to three in the illustration shown in Figure 4) is reached. To prevent premature truncation or excessive growth, we apply length normalization using a length-dependent factor α, which balances predicted CRMs probability against its length (i.e., *Normalized Score* = $(\log P(CRM)/(length)^{\alpha})$. The resulting normalized scores are then used to rank the candidate CRMs, facilitating the selection of the top-*k* chemically plausible pathways.

At this stage, it is important to note the distinction between the two kinds of top-*k* predictions utilized in our framework; one at the elementary step level and the other at the mechanism level. While the former pertains to the most probable individual elementary step predictions, the latter refers to complete mechanistic pathway obtained by applying the beam search and subsequent top-*k* ranking of different likely full CRM predictions.

**Results and Discussions**

As stated in our objectives and described in the methods section, we create a comprehensive chemical reaction mechanism dataset (ReactMech) and use them to train a generalizable graph-based DL model (DeepMech) to predict intermediates and products, beginning from the reactants as the input to the model. In the following section, we highlight the key results and its significance to different reactions.

## Performance of DeepMech on Elementary Step Prediction

As described in the methods section, the DeepMech model is designed to learn reactive bonds and classify them into distinct TMOp classes. Each TMOp encodes a template that carries the underlying mechanistic changes, such as bond breaking and bond forming. These predicted TMOps are then used to obtain the product structure (see Section C in Methods). We evaluate the performance of DeepMech on elementary step predictions by using the top-*k* accuracy, obtained by comparing the SMILES strings of the predicted products with those of the true products. Top-*k* accuracy measures the fraction of predictions of the correct product in the top-*k* predicted products ranked by the model. The top-*k* accuracies are averaged over five independent runs with distinct random train/validation/test splits. The standard deviation is provided to assess model stability. To compare the performance of our DeepMech model with the pseudo baselines, we re-trained two such models for elementary step prediction on our dataset.[39] One of these is a language-based transformer model and another a hybrid Graph2SMILES (G2S) model.[18] The transformer model considers the reaction mechanism prediction as a sequence-to-sequence task, translating the reactant SMILES into product SMILES. Specifically, we adopted the Augmented Transformer implementation as described by Joung et al.[40,18a] In contrast to the Transformer model, the G2S encodes the reactants as graphs and decodes the products in the form of SMILES sequence. In addition to the Transformer and G2S baselines, we have also compared the DeepMech performance with the most recent state-of-the-art FlowER model by the Coley group. We believe that such direct comparisons between the conceptual and implementation aspects (Table 1) together with performance on similar CRM tasks (Tables 2-4) make our contribution quite up to date. All the models were trained to predict the next intermediate or the final product.[41]

As shown in Table 2, our model consistently outperforms the G2S, Transformer, and FlowER models across all elementary step level top-*k* accuracy. The DeepMech achieves a top-1 accuracy of 98.98±0.12%, surpassing both G2S (98.00±0.14%), Transformer (93.11±0.27%) and FlowER (96.99±0.61%). The performance gap becomes more pronounced when considering higher *k* values. In many realistic scenarios, the correct product may not always be ranked first, but could appear among the top few candidates. Therefore, the use of top-*k* metrics, such as top-2 or top-3, can provide a practically more useful application of the DeepMech model. Even at top-3, our model maintains a robust accuracy of 99.46±0.08%, compared to the baselines. Statistical analysis using a paired *t*-test confirms that the improvements offered by the DeepMech model over G2S, Transformer, and FlowER are significant.[42]

**Table 2.** Elementary Step Level Top-*k* Accuracy (%) Comparison of DeepMech and Baseline Models. Reported Values are Averaged over Five Independent Runs; Standard Deviations are Shown as ±

| Model | Top-1 | Top-2 | Top-3 |
|---|---|---|---|
| DeepMech | **98.98±0.12** | **99.41±0.09** | **99.46±0.08** |
| G2S | 98.00±0.14 | 98.57±0.11 | 98.64±0.10 |
| Transformer | 93.11±0.27 | 94.94±0.20 | 95.50±0.16 |
| FlowER | 96.99±0.61 | 98.64±0.19 | 98.82±0.14 |

The relatively lower accuracy of the Transformer can be attributed to hallucination, wherein the model generates chemically implausible predictions. Typical failure modes include incorrect elementary steps and atom-level inconsistencies such as the erroneous addition or deletion of atoms, resulting in mass imbalance.[9d,18a,43] These issues are characteristic of sequence-to-sequence architectures operating on SMILES representations, where structural and chemical constraints are not explicitly enforced. Representative examples of such failures are provided in the Supporting Information Section 12. In contrast, DeepMech, which operates on TMOp with explicit mechanistic semantics, avoids such inconsistencies. Its structured representation ensures atom and charge conservation across all the mechanistic steps, thereby preventing hallucination in predictions, enabling chemically consistent outputs. Motivated by the superior predictive power and stability of our model for chemically valid elementary step prediction, we thought of considering complete CRM prediction tasks. It should be noted that a chemical reaction could have several elementary steps leading to the major product as well as those associated with the formation of byproducts.

**Complete CRM Prediction in ID Scenario**

In this section, we assess the model performance on complete CRM prediction within the ID settings. To construct the ID dataset, we use a stratified random sampling based on the number of reactions in a given reaction class, spread across all 67 reaction classes to ensure a balanced distribution. Specifically, number of samples drawn from an individual class are as follows; for reaction classes with ≤20 CRMs in them, no samples are taken, whereas those with 21–200 CRMs contributes 1–5 random samples. We sample 10% from those classes with >200 CRMs, together providing us with an ID dataset consisting of 2751 CRMs. Starting from a set of given reactants (**R**), the model predicts complete CRMs using the beam search strategy (Figure 4). It is important to note that the mechanism level top-*k* accuracy is stringently defined for this evaluation. The predicted mechanism is considered correct, only if all the elementary steps match exactly to the corresponding steps in the ground truth CRM. Since a single mismatched prediction of any elementary step renders the entire prediction incorrect, we set the accuracy of such instances to zero.[44]

**Table 3.** Mechanism Level Top-*k* Accuracy (in %) for Full CRM Prediction in the ID Setting

| Model | top-1 | top-2 | top-3 |
|---|---|---|---|
| DeepMech | **95.94±0.21** | **96.13±0.16** | **96.70±0.09** |
| G2S | 93.52±0.67 | 93.56±0.65 | 93.70±0.70 |
| Transformer | 75.27±1.28 | 75.70±1.20 | 77.60±0.92 |
| FlowER | 84.09±1.94 | 88.10±1.12 | 88.84±0.96 |

The top-*k* accuracies of our DeepMech model for CRM predictions can be compared with the other baseline models using the results provided in Table 3. Our model achieves a top-1 accuracy of 95.94±0.21%, significantly outperforming G2S (93.52±0.67%), Transformer (75.27±1.28%), and FlowER (84.09±1.94%). The performance margin widens further at higher *k* values. The paired *t*-test reveals that the performance improvements obtained with the DeepMech over the G2S are statistically significant across top-1, top-2, and top-3 accuracies, with all corresponding *p*-values below 0.05 ($p = 1.9\times10^{-3}$, $1.1\times10^{-3}$, and $1.0\times10^{-3}$, respectively). Additionally, DeepMech demonstrates better robustness, as reflected in its lower standard deviation across runs. Although the training data for the baseline models differ slightly from that used for DeepMech, all these models are evaluated on the same test set to ensure fair comparison.[39] Once again, Transformer exhibits the weakest performance, consistent with our earlier observation in the case of elementary step predictions (Table 2). The relatively lower top-*k* accuracy noted in the case of the FlowER model can be attributed to its tendency to continue generating sequences beyond the ground-truth product during the beam search, rather than terminating when the correct prediction is reached.[45] Following the promising performance of our DeepMech model, we became interested in visualizing the predictions to convey a chemically meaningful message. The predicted CRMs for different classes of representative reactions shown in Figures 5 and 6 elucidate the formation of the desired products, byproducts, and side products,[46] as well as the role of catalyst facilitating the bond-forming/breaking in multiple elementary steps of the reaction.[47]

We present a representative example in Figure 5 to illustrate the major steps involved in predicting one of the Pd-catalyzed BHA mechanisms. The DeepMech model is able to reconstruct the full catalytic cycle successfully, by predicting the key mechanistic steps such as the oxidative addition of Pd to the aryl halide ($\mathbf{R}\rightarrow\mathbf{I}_1^1$), coordination of the amine ($\mathbf{I}_1^1\rightarrow\mathbf{I}_2^1$), deprotonation ($\mathbf{I}_2^1\rightarrow\mathbf{I}_3^1$), and reductive elimination ($\mathbf{I}_3^1\rightarrow\mathbf{P}$). Importantly, the model also predicts regeneration of the active catalyst correctly. In this example, these steps are predicted as the top-1 accuracy at the mechanism level top-1.

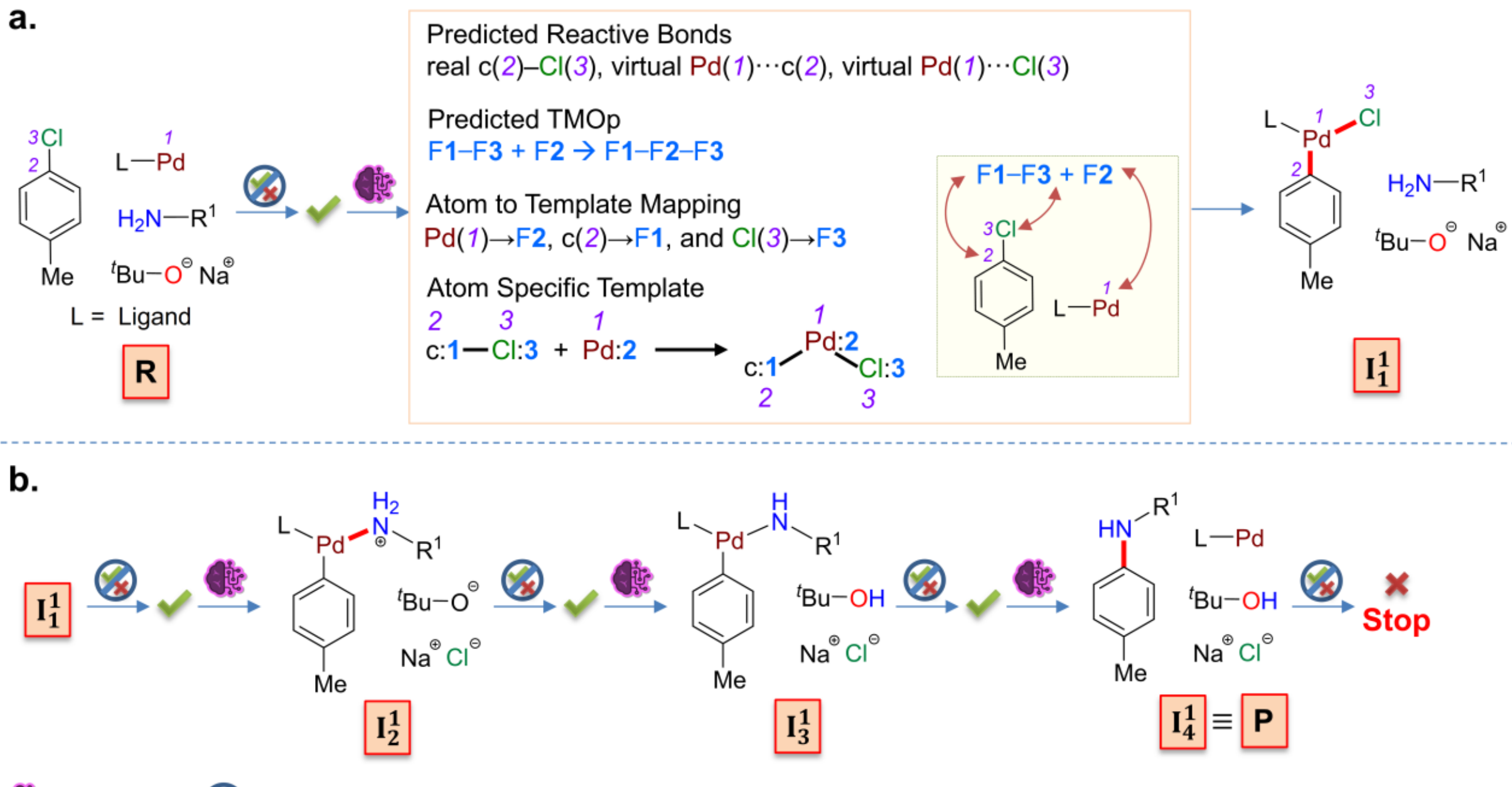

**Figure 5.** A complete CRM prediction for a representative example of the Pd-catalyzed BHA reactions between p-chlorotoulene and alkyl amine in the ID setting. (a) The process begins with evaluation by the Reaction Classifier on the given set of reactants and reagents (together denoted as **R**). When the classifier predicts **R** as reactive (✓), mechanistic inference is then carried out by the trained DeepMech model. In the first step, DeepMech identifies three reactive bonds (listed in the top row in the inset). It also predicts a TMOp for the given **R** (second row). Atom to template mapping is then performed to apply the predicted reactive atoms such as c(*2*), Cl(*3*), and Pd(*1*) to the template (third row). Substituting the atoms into the template provides the atom-specific mechanistic template (last row), which upon application to **R**, gives intermediate $\mathbf{I_1^1}$. (b) The same procedure of predicting reactive bonds, TMOp, and atom to template mapping is followed to obtain the full CRM $\mathbf{R} \rightarrow \mathbf{I_1^1} \rightarrow \mathbf{I_2^1} \rightarrow \mathbf{I_3^1} \rightarrow \mathbf{P}$.

In addition to yielding the desired products, a variety of side products are often likely to form during chemical reactions. Notably, previous models for mechanism prediction, although do well in identifying the products directly from reactants without mechanistic supervision, it remain silent on the role of catalysts and reagents, as well as in predicting side products. Identification of such species is important toward designing strategies to suppress the unwanted pathways leading to such entities. The ability of our DeepMech model in predicting plausible side products can be appreciated using a representative aromatic nucleophilic substitution reaction following an $S_NAr$ mechanism as shown in Figure 6a. The nucleophilic reactant in this case possesses two potentially reactive sites, an aliphatic amine and an aromatic amine. The mechanism level top-1 prediction by our model correctly identifies the aliphatic amine as the more nucleophilic center. An attack is initiated by the aliphatic amine to the electron-deficient aromatic ring at the aryl carbon bearing the chlorine atom, leading to an intermediate called the Meisenheimer complex. In the subsequent step, chloride is eliminated, and final deprotonation yields the desired product. Interestingly, the top-2 prediction corresponds to an analogous mechanistic pathway initiated by the aryl amine, leading to a plausible side product. While such side products are often unreported in reaction datasets, their likely anticipation could be valuable for the design of reactions and the associated optimization efforts.

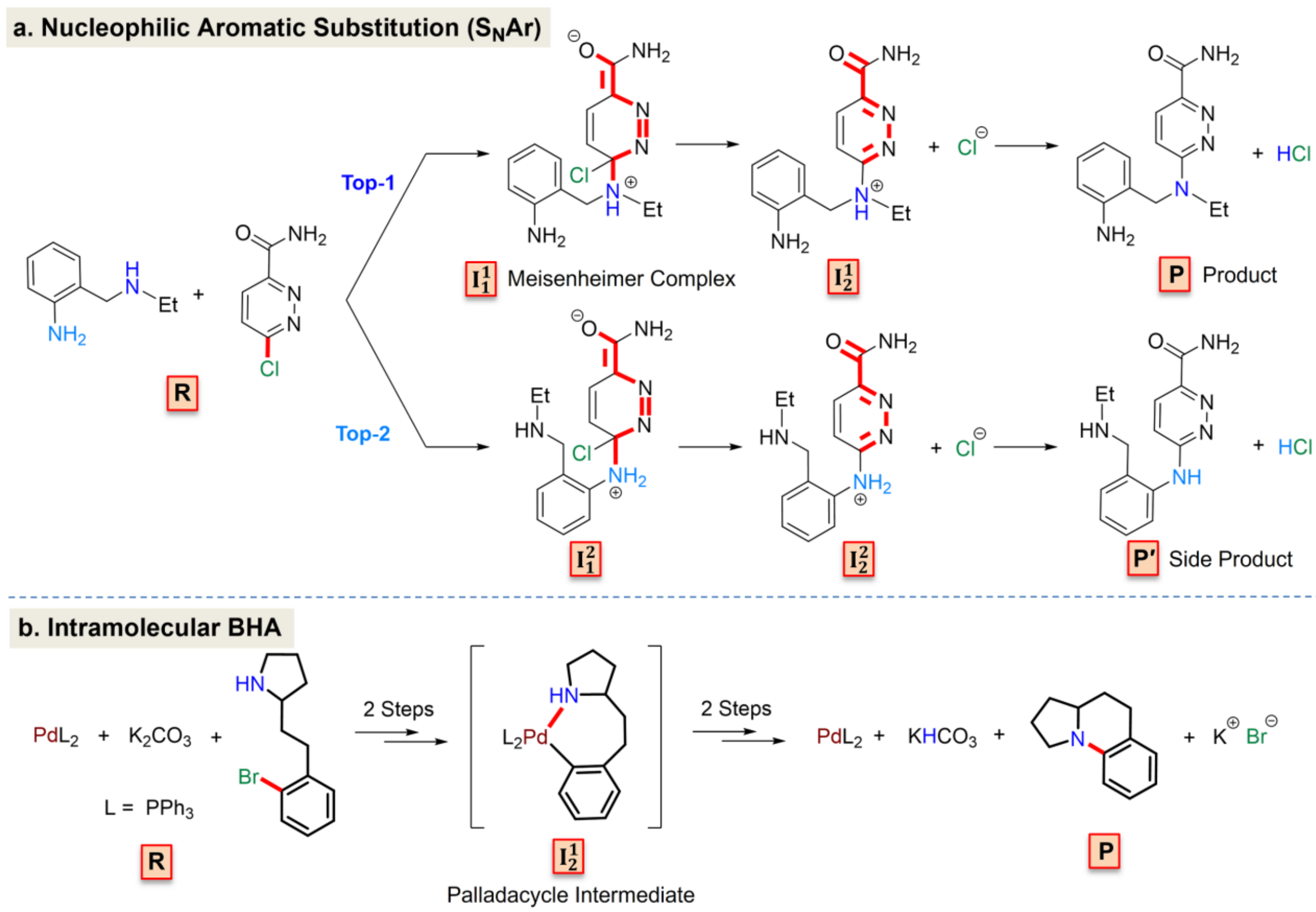

**Figure 6.** Representative examples of full CRM prediction by the DeepMech model in the case of (a) an ID nucleophilic aromatic substitution reaction, and (b) an OOD intramolecular variant of the BHA reaction.

In addition to predicting side products, another advantage of DeepMech is its ability to identify byproducts, often formed during the course of the main reaction. This capability can be traced to the explicit atom-level mechanistic reasoning built into our model. For instance, the model classifies a given step (or input) into one of the TMOp classes, which includes a template that preserves atom mapping of all atoms. No atoms are left unaccounted by the model, which help reveal potential byproducts alongside the main product, consequently imparting it with mechanistic reasoning abilities. Knowledge of byproduct formation is essential for optimizing the isolation and purification of the desired product(s) and for identifying potentially hazardous, toxic, or environmentally harmful compounds.[48] This is exemplified by using a reaction known as Appel reaction (Figure S14), wherein an alcohol gets converted to an alkyl halide by the action of triphenylphosphine and tetrahalomethane. The model accurately predicts each elementary step along with the primary substitution product, and identifies triphenylphosphine oxide as a byproduct.[49] Notably, chromatographic removal of triphenylphosphine oxide is known to be challenging because of its poor solubility in common organic solvents such as hexane or diethyl ether. Furthermore, regeneration of triphenylphosphine typically requires harsh conditions, and triphenylphosphine oxide is often discarded as waste.[50] Therefore, the ability of DeepMech in predicting byproducts as part of the overall mechanism offers significant practical value to chemists. Prior knowledge of potential byproducts could serve as a prompt for chemists to modify the choice of reagents or even redesign the reaction. Additionally, strategic removal or neutralization of byproducts at specific stages of the synthesis could as well be useful in shifting the equilibrium toward the desired product. Having obtained good accuracy in both the elementary step and full CRM predictions across a wide array of reactions in the ID setting, we considered more challenging OOD sets for evaluating the efficacy DeepMech.

**Generalization to OOD Reactions**

To evaluate generalization beyond the training distribution, we constructed an OOD dataset comprising CRMs from nucleophilic addition–elimination reactions involving amines and carbonyl compounds with different kinds of leaving groups. These compounds include acid halides, sulfonyl halides, anhydrides, haloformates, and carbamic halides. In addition, we have considered BHA reactions involving bidentate ligands and Ni catalysts, as well as its intramolecular variants. These samples are structurally different from the Pd-catalyzed ID reactions with monodentate ligands. To enhance the chemical diversity in the OOD dataset, we have also included a new mechanistic class belonging to the C–O coupling reactions (see Section 9 in the Supporting Information for details). The final OOD test set comprises of 9 mechanistically distinct classes and with a total of 3,778 CRMs. It shall be reckoned that the generalization capability of DeepMech, arises in part, from certain degree of underlying similarity in the elementary steps (TMOps) in the training set. A stricter OOD test would require reaction classes built from previously unseen TMOps. In such cases, the training set can be readily augmented with the new TMOps to enable the model to learn from the new data.[51]

A perusal of the mechanism level top-1 accuracies of the OOD reaction classes, as compiled in Table 4, suggests that the DeepMech consistently outperforms the G2S, Transformer, and FlowER across nearly all categories of reactions. For instance, the reaction class involving amine and acid halide, an accuracy of 93.55±0.72% is much superior to 60.59±4.77% obtained with the G2S and to 16.41±8.02%. Within the same class, the FlowER model performs comparatively better (84.07±2.70%) than G2S and Transformer, likely due to its underlying design principle of enforcing mass conservation. More importantly, all the baselines fail completely on reactions with anhydrides, whereas our model returns moderately good performance. Second, in the C–O coupling reaction, our model maintains a top-1 accuracy of 94.00±2.31%, outperforming all baseline models. These results together highlight the ability of our DeepMech model to generalize well to reaction classes outside the training distribution. In scenarios where the top-1 accuracy is low, slightly relaxed criteria of using higher values of $k$ led to significant improvements (see Table S8 in Supporting Information). DeepMech achieves substantial improvements in each case, notably reaching 78.33±3.11% accuracy in the intramolecular BHA class. In contrast, G2S and Transformer fail completely, and FlowER reaches a high accuracy of 70.60±5.10%. These results strongly support the superior generalizability and robustness of DeepMech in handling complex and previously unseen chemical transformations.

**Table 4.** Mechanism Level Top-1 Accuracy (%) of Full CRM Prediction across OOD Reaction Classes

| Reaction Class | | **DeepMech** | G2S | Transformer | FlowER |
|---|---|---|---|---|---|
| Amine + | acid halide | **93.55±0.72** | 60.59±4.77 | 16.41±8.02 | 84.07±2.70 |
| | sulfonyl halide | **92.42±0.78** | 7.12±4.04 | 8.46±5.49 | 76.84±6.11 |
| | anhydride | **64.51±0.26** | 0.00±0.00 | 0.07±0.17 | 0.00±0.00 |
| | haloformate | **92.05±2.17** | 31.48±9.19 | 20.39±8.98 | 70.56±0.56 |
| | carbamic halides | **92.42±2.83** | 53.03±6.06 | 19.70±8.09 | 90.91±2.63 |
| C–O coupling | | **94.00±2.31** | 16.00±4.00 | 0.00±0.00 | 0.00±0.00 |
| BHA | intramolecular | **78.33±3.11** | 0.00±0.00 | 0.0±0.00 | 70.60±5.10 |
| | with bidentate ligand | **99.90±0.22** | 0.00±0.00 | 0.00±0.00 | 57.00±1.98 |
| | Ni-catalyzed | **100.00±0.00** | 0.00±0.00 | 0.00±0.00 | 72.38±2.81 |

Another interesting application of DeepMech in predicting the elementary steps in a Pd-catalyzed intramolecular BHA reaction is shown in Figure 6b, in which an aryl halide and amine moieties tethered within the same substrate leading to a cyclic product is considered. The mechanism is correctly predicted by the model, which begins with the oxidative addition of Pd to the $C_{aryl}$–Br bond, followed by intramolecular coordination of the pendant amine to the Pd center to form a palladacycle intermediate. Importantly, such palladacycles were not present in the training set. In the ensuing mechanistic steps, both deprotonation and reductive elimination is predicted in such a way that it leads to the desired benzo-fused indolizidine product, an important structural motif found in naturally occurring alkaloids[52] and regenerates the active catalyst.

These examples collectively underscore the effectiveness of DeepMech in generalizing well beyond the reaction classes in the training set.

Consistently good performance of the DeepMech model on both ID and OOD deployments prompted us to wonder about possible factors contributing to its broad-spectrum capabilities. We consider the following as probable origin of the robustness across different reaction classes and absence of hallucination noted in the case of Transformer-based models. Since the DeepMech learns to identify reactive bonds, it remains robust to highly diverse chemical reaction datasets. This means that, when presented with novel molecules exhibiting similar reactivity, DeepMech can accurately detect the reactive bonds and the associated TMOp (Template, Operation, $\Delta$H, and $\Delta$q), hence predict the mechanistic steps. Suppose, a complex natural product structure bearing an aryl chloride (or aryl halide, in general) moiety, among other functional groups, is the input candidate along with a suitable PdL as the catalyst, the model would be able to correctly predict an oxidative addition product/intermediate at the Ar–Cl bond. This is because the model is trained to learn the underlying reactivity pattern, and can therefore faithfully generate the correct product/intermediate in OOD settings as well. In contrast, G2S or Transformer-based models would have to generate the entire molecule, including the complex scaffold of the natural product that these models were not trained with explicitly. As a result, such models are likely to fail in OOD scenarios. Similarly, integration of DeepMech with TMOp makes it less susceptible to hallucination. DeepMech would not predict a new atom that doesn't exist in the input molecular graph, unlike the other pseudo baseline models, which could predict a Ni intermediate, even if the input belongs to a Pd-catalyzed BHA reaction. Another probable cause of poorer OOD performance of the pseudo baseline models might be due to incorrect prediction of certain elementary steps due to hallucination of atom(s) and/or invalid SMILES generation. It should be noted that one incorrect elementary step prediction in beam search would trigger a cascade of generation of wrong intermediates/product rendering the full CRM invalid.

As with any other ML-based model, there are certain aspects with the accuracy of DeepMech while deploying it for full CRM predictions. A few such observations are worth considering at this point. First, DeepMech performs exceedingly well in elementary step as well as multi-step mechanism predictions with ID examples. A comparatively lower performance with the multi-step OOD case (Table 4) could arise from issues with the beam search and/or reaction classifier.[53] For instance, a full CRM prediction task is prone more errors as an incorrect prediction in any one elementary step during the beam search would render the whole mechanism invalid, even amounting to a zero in top-1 accuracy. Since the beam search classifier contains stopping as an outcome, a premature termination or prolongation of a predicted sequence beyond a chemically meaningful product could lead to unexpected errors. A potential limitation of our approach is the reliance on a predefined set of TMOps, which could constrain the prediction of truly novel reactivity. However, we argue this is also a strength, as it grounds the model in chemical reality and prevents the unconstrained hallucination common in other architectures. Furthermore, our framework is modular, allowing for the future expansion of the TMOp library as new elementary reactions are characterized.[51] Despite these challenges in full CRM predictions, DeepMech remains robust in elementary step prediction, even in with OOD instances of similar reactivity. Promising new results in a relatively difficult task of reaction mechanism prediction demonstrate the ability of DeepMech in effective abstraction of reaction mechanisms. To probe into how DeepMech generalizes well across diverse reaction classes and to it’s potential interpretability, we have used graph attention visualization. In particular, in the following section, the attention on the reacting entities is analyzed toward developing an improved understanding of learnable aspects of DeepMech, when presented it with reaction mechanism datasets.

**Attention Visualization**

Given that DeepMech is constructed upon both atom- and bond-level attention mechanisms, as described in Figure 3, we have probed the attentive constituents to gather mechanistic insights and interpretability.[54] Interestingly, the atoms and bonds receiving high attention are found to be the ones expected to be the chemically reactive centers. This concordance can be appreciated with the help of three representative elementary steps as shown in Figure 7. This analysis highlights the potential of attention-based interpretation of predictions offered by the model.

In Figure 7a, we present an oxidative addition step typical of a BHA reaction. The atom-level attention convincingly encompassed the primary reactive atoms (Pd and Cl) involved in this specific step, as well as secondary reactive atoms such as the nitrogen of the coupling partner. This suggests the model ability to recognize and weigh not only the immediate reaction site but also downstream reaction participants, thereby exhibiting enhanced interpretability spanning multiple stages. Importantly, adjacent atoms to the reactive centers (P and α-C of the ligand entity) also received moderate attention, which may reflect the influence of electronic and steric substituent effects, highlighting the model sensitivity to subtle, but chemically relevant, features. Additionally, the bond-level attention correctly focused on the reactive bonds in this elementary step, primarily highlighting the C–Cl bond undergoing cleavage and the Pd–C and Pd–Cl bonds being formed. Impressively, these bonds ranked among the top five most attentive, emphasizing their pivotal influence in this transformation. This marks the first instance, to the best of our knowledge, in which both real and virtual bonds have been jointly considered and substantially emphasized within an attention-based framework, representing a noteworthy advancement in the modeling of chemical reactivity (see Section 8 in Supporting Information for more details).

Figure 7b demonstrates another compelling scenario involving a nucleophilic attack by phosphorus on a bromine atom of a symmetric tetrabromomethane. The full mechanism of this reaction, with several elementary steps, is provided in Figure S12. The model effectively identified all the Br atoms as highly attentive, reflecting their chemical equivalency and potential reactivity. Additionally, atoms such as the hydroxyl oxygen in the substrate, although not reactive in the triphenylphosphine activation step, still garnered attention because of their involvement in subsequent steps. Furthermore, virtual bonds connecting phosphorus to bromine received significantly elevated attention scores, indicating the model ability to identify chemically reactive bonds. Another interesting example depicted in Figure 7c involves the deprotonation of a tertiary alcohol facilitated by sodium hydride, an elementary step within a nucleophilic substitution proceeding via an addition-elimination mechanism. Notably, the model assigns the highest atom-level attention to the hydride ion, underscoring its ability to accurately capture the reactivity of small, yet mechanistically critical, species. Complementing this, the virtual bond connecting the hydride ion to the oxygen atom of the alcohol receives the highest bond-level attention, further validating the model effectiveness to highlight key reactive bonds that define the elementary step. These layered attention mechanisms (operating at both bond and atom levels) provide an enhanced interpretability of DeepMech. Collectively, these findings serve as a convincing demonstration of the utility of our DeepMech model in identifying potential reactive bonds and atoms.

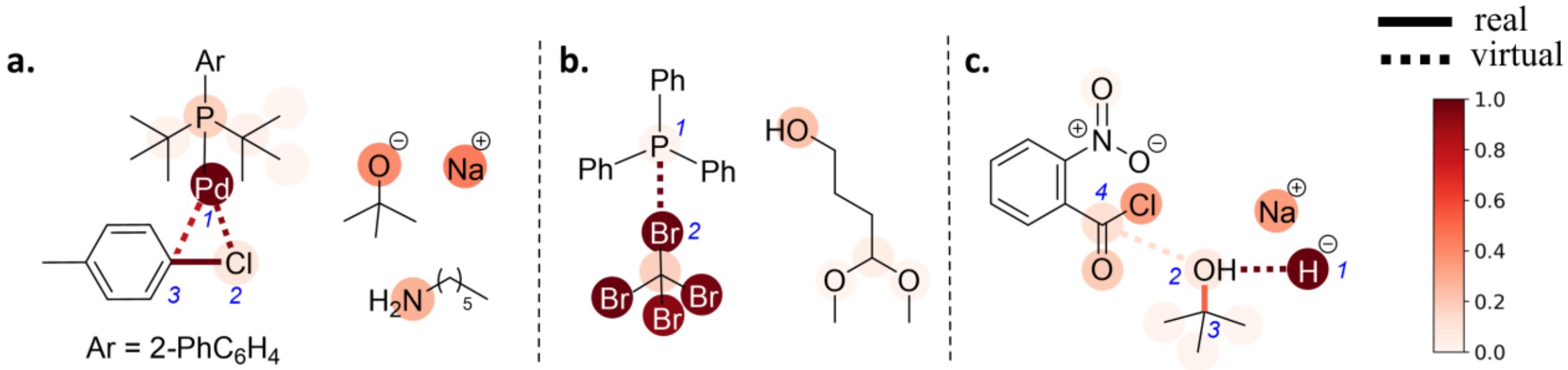

**Figure 7.** Visualization of atom and bond attention in three representative elementary reaction steps: (a) oxidative addition in a BHA reaction, (b) carbanion formation in an Appel reaction, (c) proton abstraction step in a nucleophilic substitution reaction through an addition-elimination mechanism. Atom attentions (top-10) are shown directly on molecular structures with colored circles, while selected bonds with higher attention values are highlighted through color coding and increased line thickness for clarity. Higher color intensity indicates greater attention.

**Application of DeepMech to Prebiotic Chemistry**

To examine the applicability of our DeepMech model, we apply it to reaction mechanisms relevant to prebiotic chemistry, which concern the abiotic synthesis of biologically essential molecules from simple primordial substrates. Elucidating the mechanistic pathways that lead to the formation of the basic building blocks, such as amino acids, sugars, and other biomolecular precursors from basic prebiotic inputs, such as $H_2O$, $N_2$, HCN, $NH_3$, and $CH_4$, is of fundamental importance to understanding the chemical origins of life. To systematically explore such chemical space, we followed the protocol outlined in the section 'Dataset Preparation' to generate a large-scale dataset of 30,787 elementary steps.[55] This dataset, which we refer to as PrebioMech, represents a diverse and chemically reasonable set of reaction mechanisms in line with prebiotic conditions (see Section 10 in Supporting Information for further details). Note that the ReactMech and PrebioMech datasets were trained and evaluated separately because they originate from two fundamentally different data distributions. Specifically, ReactMech comprises reactions that are experimentally validated and represents chemical transformations commonly observed under standard laboratory conditions. In contrast, PrebioMech includes prebiotic and non-traditional reaction mechanisms, many of them appear chemically unusual and not amenable under standard reaction conditions.[56]

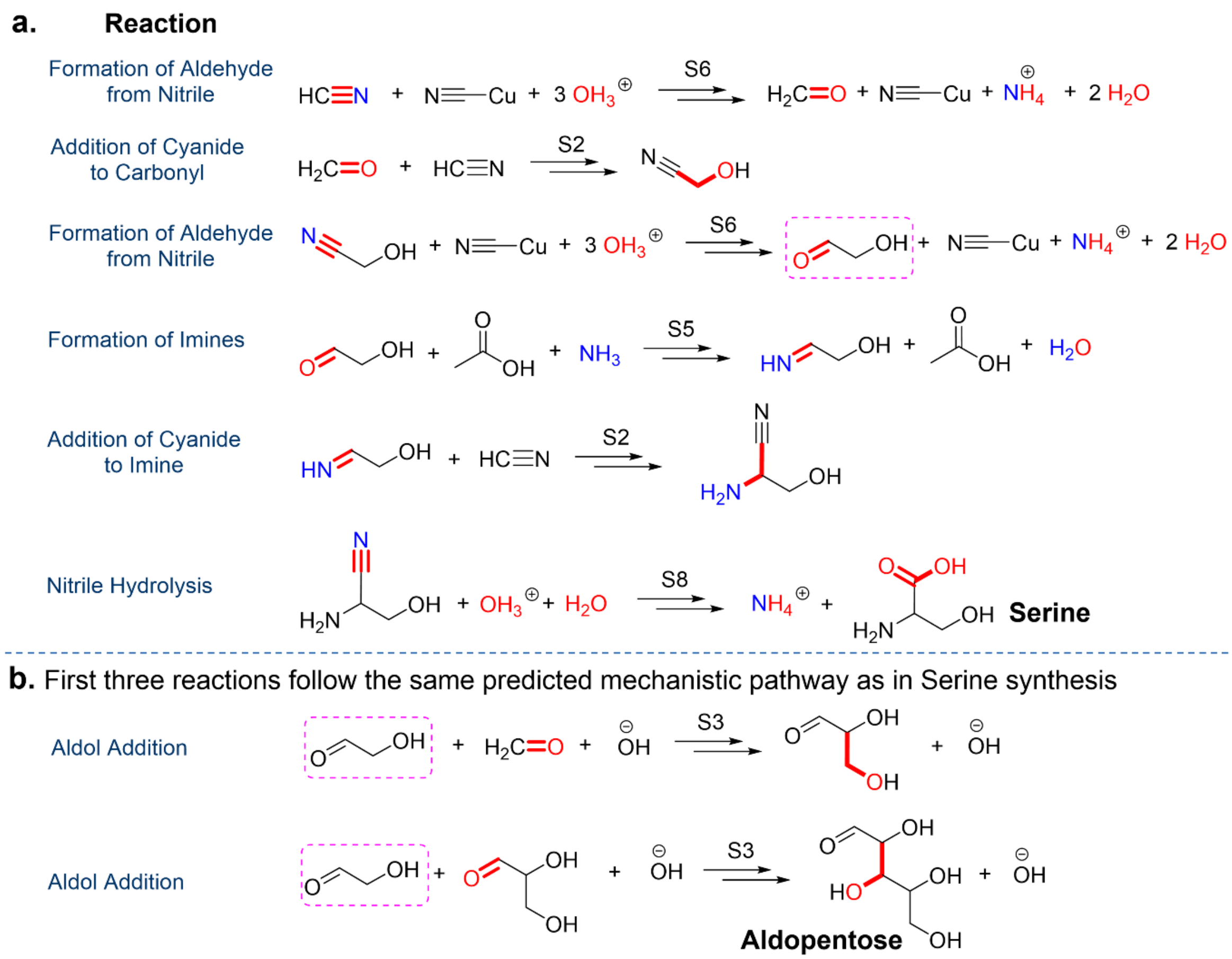

**Figure 8.** Predicted mechanistic steps for the synthesis of (a) Serine (6 reactions) and (b) Aldopentose (5 reactions) using DeepMech starting from prebiotic compounds. Here, number of elementary steps in each transformation is denoted as Sn where n=1,2,3, …,n.

Using this dataset, we re-train the DeepMech model and evaluate its ability to predict mechanistic pathways leading to the formation of amino acids and other biologically relevant molecules. Consistent with its performance on the ReactMech dataset, DeepMech attains a top-1 accuracy of 96.03% for elementary step prediction in this dataset.[57] Remarkably, the model successfully reconstructs complete mechanistic routes for several amino acids. Figure 8a illustrates the complete reaction network (see Section 10 in Supporting Information for the associated predicted mechanisms) that leads to the formation of serine, a key naturally occurring α-amino acid. The mechanistic prediction of serine formation pathway uses formaldehyde, which in turn, is obtained from hydrogen cyanide (see first row in Figure 8a). DeepMech reconstructs al the likely elementary steps. In the second row, the resulting formaldehyde reacts with HCN via two steps: a protonation followed by nucleophilic attack to yield an α-cyano alcohol. In the third row, this alcohol undergoes a series of mechanistic steps analogous to those in the initial formaldehyde formation, leading to the formation of an α-hydroxy aldehyde. In the fourth row, ammonia reacts with the α-hydroxy aldehyde to form the corresponding imine. Subsequently, in the fifth row, a second equivalent of hydrogen cyanide is added to the imine through a mechanism similar to that described in row two. Finally, the resulting 2-amino-3-hydroxypropanenitrile undergoes nitrile hydrolysis, a sequence comprising nine elementary steps, culminating in the formation of the amino acid serine (shown in the sixth row). The model accurately predicts all 29 elementary steps, validating the effectiveness of our TMOp based approach for reconstructing complete reaction mechanisms. Here, the inclusion of CuCN (as shown in rows 1 and 3 in Figure 8a) reflects its indirect but essential photo-catalytic role. Specifically, CuCN participates in a photochemical process and subsequently reacts with HCN, producing two electrons and two protons that facilitate the overall transformation.[58] We have approximated its effect by introducing two protons and treating CuCN as a spectator species in the simplified mechanism. While this represents a coarse grain abstraction, it serves as a proof of concept to demonstrate that DeepMech can be customized to handle reactions relevant to prebiotic chemistry under some simplified assumptions.

In addition to serine, DeepMech successfully predicts the formation of an important biologically relevant target molecule, specifically an aldopentose, as illustrated in Figure 8b. The initial three reaction steps leading to the formation of aldopentose are the same as those involved in serine biosynthesis. The resulting α-hydroxy aldehyde subsequently undergoes a base-catalyzed aldol addition with formaldehyde to yield an α,β-dihydroxy aldehyde. This aldehyde then participates in a second aldol addition with another molecule of α-hydroxy aldehyde, ultimately producing the aldopentose sugar. These results underscore the versatility and generalizability of the DeepMech framework that goes beyond the conventional reactions. More broadly, this demonstrates that our TMOp-based approach can be extended to other important areas such as astrochemistry, biochemical transformations, metabolic pathways, and petrochemical processes. As a future direction, we could continue integrating DeepMech with experiment or ab initio calculations to refine its fidelity to complex mechanistic problems.

**Conclusions**

We present DeepMech, a graph-based deep learning model for complete chemical reaction mechanism (CRM) prediction and ReactMech dataset containing mechanistic steps of a wide array of chemical reactions. Unlike sequence-to-sequence models that generate products of reactions by using a character-by-character approach, making them vulnerable to hallucinations, our DeepMech is designed to learn reactive bonds and template of mechanistic operations (TMOp). The integration of subgraph isomorphism, a reaction classifier as a stopping criterion, and a tailored beam search approach together is found to be effective in accurate identification of full and chemically plausible mechanistic pathways. Trained on the meticulously curated

ReactMech dataset containing reactions from USPTO and transition metal catalysis, DeepMech achieves a high accuracy of 98.98±0.12% in elementary step predictions and 95.94±0.21% with in-distribution CRM predictions, outperforming pseudo baselines such as Graph2SMILES, Transformers and FlowER. The generalizability of the DeepMech model became evident from the robust performance with the out-of-distribution CRM prediction tasks, achieving superior accuracies than the baseline models. The core design components of DeepMech is built around the identification of reactive atoms and bonds involved in each elementary step, due to which the model offers mechanistic interpretability and insights consistent with chemical intuition. We could further showcase the broader applicability of DeepMech by successfully reconstructing multistep mechanisms relevant to prebiotic chemistry, starting from simpler precursors such as nitrogen, ammonia, methane and others, to important biological building blocks amino acid/sugars. In summary, DeepMech is a reliable, generalizable, and interpretable framework for CRM prediction that integrates data-driven modeling with mechanistic principles, which can be potentially extended to advanced predictive chemistry, reaction development and mechanism validation.

**ASSOCIATED CONTENT**

**Corresponding Authors**

**Raghavan B. Sunoj**: Department of Chemistry and Centre for Machine Intelligence and Data Science, Indian Institute of Technology Bombay, Mumbai, Maharashtra, 400076, India; https://orcid.org/0000-0002-6484-2878; Email: sunoj@chem.iitb.ac.in

**Mayank Baranwal**: Department of Systems and Control Engineering, Indian Institute of Technology Bombay, Powai, Mumbai 400076, India; https://orcid.org/0000-0001-9354-2826; Email: mbaranwal@iitb.ac.in

**Authors**

**Manajit Das**: Department of Chemistry, Indian Institute of Technology Bombay, Powai, Mumbai 400076, India; https://orcid.org/0000-0001-7709-8809

**Ajnabiul Hoque**: Department of Chemistry, Indian Institute of Technology Bombay, Powai, Mumbai 400076, India; https://orcid.org/0000-0001-9807-3061

**Supporting Information**

The details of DeepMech model, data analysis, training details (including loss, hyperparameters, subgraph isomorphism and reaction classifier), full CRM prediction (ID and OOD), prebiotic predictions, ablation studies, failure cases in baseline models, and top-k accuracy calculation at the mechanism level are included.

**Acknowledgements**

We are thankful to Institution of Eminence (IoE) Data and Information Science computing facility for generous computational resources. M.D. acknowledges the Prime Minister's Research Fellowship.

**Notes**

The authors declare no competing financial interest.

**Code and data availability**

The datasets and source code utilized for training the models are accessible via the following link: https://github.com/alhqlearn/DeepMech